%% file: InclusivePhotons.tex
\documentclass[twocolumn,showpacs,aps,prd]{revtex4-1}

\usepackage{preprintcover}
\input{cover.tex}

\usepackage{graphicx} 

\usepackage{atlasphysics} 

\graphicspath{}

\usepackage{amsmath}
\usepackage{atlasphysics}
\usepackage{url}
\usepackage{subfigure}
\usepackage{epstopdf}
\usepackage{multirow}
\usepackage{xspace}
\usepackage{atlasphysics}
\usepackage{rotating} 
\usepackage{longtable}
\usepackage{multirow}
\usepackage{url}
\usepackage{hyperref}
\hypersetup{
  colorlinks=true,
  linkcolor=blue,
  citecolor=blue,
  urlcolor=blue
}

\newcommand{\ETg}{\ensuremath{E_{\mathrm T}^{\gamma}}}
\newcommand{\ETAg}{\ensuremath{\eta^{\gamma}}}
\newcommand{\ETiso}{\ensuremath{E_{\mathrm T}^{\mathrm{iso}}}}

\date{\today}

\begin{document}

\title{\boldmath Measurement of the inclusive isolated prompt photons cross section 
  in $pp$ collisions at $\sqrt{s}= 7$~TeV with the ATLAS
  detector using 4.6 fb$^{-1}$ }
\collaboration{The ATLAS Collaboration}

%
%

\begin{abstract}
 A measurement of the cross section for the
 production of isolated prompt photons in $pp$ collisions at a
 center-of-mass energy $\sqrt{s} = 7$~TeV is presented.  
 The results are based on an
 integrated luminosity of 4.6~\ifb\ collected with the ATLAS detector
 at the LHC.
 The cross section is measured as a function of photon pseudorapidity $\ETAg$ and transverse energy $\ETg$ in the
 kinematic range $100\leq\ETg<1000$~GeV and in the regions $|\ETAg|<1.37$ and $1.52\leq|\ETAg|<2.37$.
The results are compared to 
leading-order parton-shower Monte Carlo models
and next-to-leading-order perturbative QCD
calculations.  Next-to-leading-order perturbative QCD calculations 
agree well with the measured cross sections as a function of $\ETg$ and $\ETAg$.
\end{abstract}

\pacs{13.25.Hw, 12.15.Hh, 11.30.Er}

\maketitle


\input{definitions}



\newpage

\section{Introduction}
\label{sec:Introduction}
\input{introduction}

\section{The ATLAS detector}
\label{sec:Detector}
\input{detector}

\section{Data and simulated samples}
\label{sec:Samples}
\input{data_samples}

\section{Photon selection}
\label{sec:PhotonID}
\input{photon_reconstruction}

\section{Background estimation and signal extraction}
\label{sec:Purity}
\input{purity}

\section{Residual background}
\label{sec:CrossSection}
\input{unfolding}

\section{Cross-section measurement}
\label{sec:efficiency}
\input{efficiency}

\section{Systematic uncertainties}
\label{sec:uncertainties}
\input{systematic_Uncertainties}


\section{Theoretical predictions}
\label{sec:Theory}
\input{theory}

\section{Results}
\label{sec:results}
\input{theorycomparison}

\section{Conclusion}
\label{sec:Conclusion}
\input{conclusion}

\input{tables}

\section*{Acknowledgements}

We thank CERN for the very successful operation of the LHC, as well as the
support staff from our institutions without whom ATLAS could not be
operated efficiently.

We acknowledge the support of ANPCyT, Argentina; YerPhI, Armenia; ARC,
Australia; BMWF and FWF, Austria; ANAS, Azerbaijan; SSTC, Belarus; CNPq and FAPESP,
Brazil; NSERC, NRC and CFI, Canada; CERN; CONICYT, Chile; CAS, MOST and NSFC,
China; COLCIENCIAS, Colombia; MSMT CR, MPO CR and VSC CR, Czech Republic;
DNRF, DNSRC and Lundbeck Foundation, Denmark; EPLANET, ERC and NSRF, European Union;
IN2P3-CNRS, CEA-DSM/IRFU, France; GNSF, Georgia; BMBF, DFG, HGF, MPG and AvH
Foundation, Germany; GSRT and NSRF, Greece; ISF, MINERVA, GIF, DIP and Benoziyo Center,
Israel; INFN, Italy; MEXT and JSPS, Japan; CNRST, Morocco; FOM and NWO,
Netherlands; BRF and RCN, Norway; MNiSW, Poland; GRICES and FCT, Portugal; MERYS
(MECTS), Romania; MES of Russia and ROSATOM, Russian Federation; JINR; MSTD,
Serbia; MSSR, Slovakia; ARRS and MIZ\v{S}, Slovenia; DST/NRF, South Africa;
MICINN, Spain; SRC and Wallenberg Foundation, Sweden; SER, SNSF and Cantons of
Bern and Geneva, Switzerland; NSC, Taiwan; TAEK, Turkey; STFC, the Royal
Society and Leverhulme Trust, United Kingdom; DOE and NSF, United States of
America.

The crucial computing support from all WLCG partners is acknowledged
gratefully, in particular from CERN and the ATLAS Tier-1 facilities at
TRIUMF (Canada), NDGF (Denmark, Norway, Sweden), CC-IN2P3 (France),
KIT/GridKA (Germany), INFN-CNAF (Italy), NL-T1 (Netherlands), PIC (Spain),
ASGC (Taiwan), RAL (UK) and BNL (USA) and in the Tier-2 facilities
worldwide.

\newpage
\clearpage
\bibliography{InclusivePhotons}



\onecolumngrid
\clearpage
\input{atlas_authlist}

\end{document}

%% file: cover.tex
\PreprintCoverPaperTitle{\boldmath
Measurement of the inclusive isolated prompt photon cross section
in $pp$ collisions at $\sqrt{s} = 7\TeV$ with the ATLAS detector \\ using 4.6~fb$^{-1}$}  
\PreprintIdNumber{CERN-PH-EP-2013-164}  
\PreprintCoverAbstract{ A measurement of the cross section for the
 production of isolated prompt photons in $pp$ collisions at a
 center-of-mass energy $\sqrt{s} = 7$~TeV is presented.
 The results are based on an
 integrated luminosity of 4.6~\ifb\ collected with the ATLAS detector
 at the LHC.
 The cross section is measured as a function of photon pseudorapidity $\ETAg$ and transverse energy $\ETg$ in the
 kinematic range $100\leq\ETg<1000$~GeV and in the regions $|\ETAg|<1.37$ and $1.52\leq|\ETAg|<2.37$.
The results are compared to
leading-order parton-shower Monte Carlo models
and next-to-leading-order perturbative QCD
calculations.  Next-to-leading-order perturbative QCD calculations
agree well with the measured cross sections as a function of $\ETg$ and $\ETAg$.}  
\PreprintJournalName{Phys. Rev. D}  

%% file: definitions.tex
%
\newcommand{\Rhad}    {\ensuremath{R_{\rm had}}}
\newcommand{\Rhadone} {\ensuremath{R_{\rm had_{1}}}}

\newcommand{\Reta}    {\ensuremath{R_{\eta}}}
\newcommand{\Rphi}    {\ensuremath{R_{\phi}}}
\newcommand{\wetatwo} {\ensuremath{w_{2}}}

\newcommand{\Fside}   {\ensuremath{F_{\rm side}}}
\newcommand{\wthree}  {\ensuremath{w_{s\,3}}}
\newcommand{\wtot}    {\ensuremath{w_{s\,{\rm tot}}}}
\newcommand{\DeltaE}  {\ensuremath{\Delta{}E}}
\newcommand{\Eratio}  {\ensuremath{E_{\rm ratio}}}
\newcommand{\Emaxtwor}{\ensuremath{R_{\rm max\,1}}}

\newcommand{\mo}      {\ensuremath{^{-1}}}

\newcommand{\Etg}     {\ensuremath{E_{\rm T}^{\gamma}}}
\newcommand{\etag}     {\ensuremath{\eta^{\gamma}}}

\newcommand{\effID}{\hbox{$\varepsilon^{{\rm offl}}$}}
\newcommand{\effReco}{\hbox{$\varepsilon^{{\rm reco}}$}}

\newcommand{\EtCone}{\et^{\rm iso}}

\newcommand{\obs}{^\mathrm{obs}}
\newcommand{\sig}{^\mathrm{sig}}
\newcommand{\bkg}{^\mathrm{bkg}}
\newcommand{\Ns}{N\sig}
\newcommand{\Nb}{N\bkg}
\newcommand{\N}{N\obs}

\newcommand{\pdf}{\Pi}
\newcommand{\ph}{_\gamma}
\newcommand{\MC}{_\mathrm{MC}}
\newcommand{\data}{_\mathrm{data}}

\newcommand{\etaM}{|\eta_\mathrm{S2}|}

\newcommand{\Equation}[2]{\begin{equation} #2 \label{#1} \end{equation} }

%% file: introduction.tex
%
%
Prompt photon production at hadron colliders allows
tests of perturbative QCD predictions~\cite{QCD_promptphoton_NLO}.
The measurement is sensitive to the gluon content of the
proton through the $qg~\rightarrow~q\gamma$ process, which 
dominates the prompt photon production cross section at the LHC, and 
can be used to constrain parton distribution
functions (PDFs)~\cite{promptphoton_and_gluon_pdf}.

%
%
Recent measurements of the production cross section of isolated prompt photons
have been performed by ATLAS~\cite{Aad:2010sp,Aad:2011tw} and 
CMS~\cite{CMSpreviouspaper,PhysRevLett.106.082001} using $pp$ collision data at $\sqrt{s}=7$~TeV at the LHC.
Earlier measurements were made by CDF and D0 using $p\bar{p}$  collisions collected
at $\sqrt{s}=1.8$~TeV  and  $\sqrt{s}=1.96$~TeV at the Tevatron collider \cite{Acosta:2002ya,Acosta:2004bg,CDF_promptphoton,D0_promptphoton}. 
Also, similar measurements were made at the $Sp\bar{p}S$ collider \cite{SppbarS,SppbarS2}.

%
%
In this paper, the production cross section of isolated prompt photons 
is measured in the transverse energy ($\ETg$) range between 100~GeV and 1~TeV,
extending the result of the previous ATLAS measurement, which covered
the range between 45~GeV and 400~GeV~\cite{Aad:2011tw}. 
The differential cross section as a function of $\ETg$ 
is measured in the
pseudorapidity~\footnote{ATLAS
  uses a right-handed coordinate system with its origin at the nominal
  interaction point (IP) in the center of the detector, and the
  $z$-axis along the beam line.  The $x$-axis points from the IP to
  the center of the LHC ring, and the $y$-axis points upwards.
  Cylindrical coordinates $(r,\phi)$ are used in the transverse plane,
  $\phi$ being the azimuthal angle around the beam line.  Observables
  labeled ``transverse'' are projected into the $x$--$y$ plane.
  The pseudorapidity is defined in terms of the polar angle $\theta$ as
  $\eta=-\ln\tan(\theta/2)$.} 
range  $|\ETAg|<1.37$ (the barrel region) and $1.52\leq
|\ETAg| < 2.37$ (the end-cap region). 
Photon reconstruction in these pseudorapidity regions has a  high
efficiency and a low background rate.
The differential cross section is also studied as a 
function of $\ETAg$ for $\ETg>100$~GeV.
The data sample corresponds to
an integrated
luminosity of $4.64\pm 0.08~\ifb$ \cite{Aad:2013ucp}; thus
this analysis uses a data set more than two orders of magnitude larger than that
used in the previous measurement~\cite{Aad:2011tw}.


%
%
In the following, all photons produced in $pp$ collisions and that are not
secondaries to hadron decays are considered as ``prompt''. They include
``direct'' photons, which originate from the hard processes calculable in perturbative QCD, 
and ``fragmentation'' photons, which are the result of the
 fragmentation of a colored high-$\pt$ parton~\cite{jetphox1,PhysRevD.73.094007}.
Photons are considered ``isolated'' if the transverse energy ($\ETiso$) within
a cone of radius $\Delta R=\sqrt{(\Delta \eta)^2 + (\Delta \phi)^2}=0.4$
centered around the photon in the pseudorapidity and azimuthal angle ($\phi$)
is smaller than 7~GeV. 
In next-to-leading-order (NLO) parton-level theoretical calculations, 
$\ETiso$ is calculated from all partons within the cone, while in the 
leading-order (LO) parton-shower Monte Carlo (MC) simulations, $\ETiso$ is calculated from all
the generated particles (except muons and neutrinos) inside the cone.
Experimentally, $\ETiso$ is calculated from the energy
deposited in the calorimeters in a $\Delta R=0.4$ cone around the photon
candidate, corrected for effects associated with the energy of the
photon candidate itself, the underlying event, and the additional $pp$
interactions in the same bunch crossing (pileup)~\cite{:2012je}.
The main background for the prompt photons consists of photons from decays of light neutral 
mesons such as the $\pi^{0}$ or $\eta$.

%% file: detector.tex
ATLAS~\cite{ATLAS_detector} is a multi-purpose detector with a forward-backward 
symmetric cylindrical geometry and nearly 4\pi\ coverage in solid angle. 
The most relevant subdetectors for the present 
analysis are the inner tracking detector (ID) and the calorimeters.
 
The ID consists of a silicon pixel detector and a silicon microstrip
detector covering the pseudorapidity 
range $|\eta|<2.5$,
and a straw-tube transition radiation tracker covering $|\eta|<2.0$. 
It is immersed in a 2 T magnetic field provided by a superconducting
solenoid.
The ID allows efficient reconstruction of converted photons if
the conversion occurs at a radius of less than $0.80$~m.

The electromagnetic calorimeter (ECAL) is a lead/liquid-argon (LAr)
sampling calo\-ri\-me\-ter providing coverage for $|\eta|< 3.2$.
It consists of a barrel section ($|\eta| < 1.475$) and two
end-caps ($1.375<|\eta|< 3.2$). The central region ($|\eta|<2.5$)
is segmented into three layers in shower depth.
The first (inner) layer, covering $|\eta| < 1.4$ in the barrel and 
$1.5<|\eta|< 2.4$ in the end-caps, has a high $\eta$
granularity (between 0.003 and 0.006 depending on $\eta$),
which can be used to provide event-by-event discrimination
between single-photon showers and two overlapping
showers such as may be produced by $\pi^0$ decay. 
The second layer, which collects most of the energy deposited 
in the calorimeter by the photon shower, has a cell 
granularity of 0.025$\times$0.025 in $\eta \times \phi$. 
The third layer is used to correct high-energy showers for leakage.
In front of the ECAL a thin presampler layer, covering the
pseudorapidity interval $|\eta|< 1.8$, is used to correct for
energy loss before the ECAL.

The hadronic calorimeter (HCAL), surrounding the
ECAL, consists of an iron/scintillator-tile calorimeter
in the range $|\eta|< 1.7$, and two copper/LAr calorimeters
spanning $1.5 <|\eta|< 3.2$. The ECAL and HCAL acceptance is extended
by two copper/LAr forward calorimeters (using copper and tungsten
as absorbers) up to $|\eta|=4.9$.

A three-level trigger system is used to select events containing
photon candidates. 
The first level (level-1) is implemented in hardware and is based on
towers with a coarser granularity ($0.1\times 0.1$ in $\eta\times\phi$)
than that of the ECAL.
They are used to search for electromagnetic deposits 
in $\eta \times \phi$ regions of $2\times 1$ and $1\times 2$ towers, 
within a fixed window of size $2\times 2$ and with an 
$\ETg$ above a programmable threshold. 
The algorithms of the second and third level 
triggers (collectively referred to as the {\em high-level trigger})
are implemented in software. The
high-level trigger exploits the full granularity and precision of the
calorimeter to refine the level-1 trigger selection, based on improved
energy resolution and detailed information on energy deposition in the
calorimeter cells.

%% file: data_samples.tex
%
\subsection{Collision Data selection}
%
%
The measurement presented here is based on proton-proton collision
data collected at a center-of-mass energy of $\sqrt{s}=7$~TeV with the
ATLAS detector at the LHC in 2011.
Only events where both the calorimeter and the ID are
fully operational and that have good data quality are used. 
Events are triggered using a high-level photon trigger,
with a nominal $\ETg$ threshold of 80~GeV.
The trigger selection criteria for the fraction and profile of the
energy measured in the various layers of the calorimeters are looser
than the photon identification criteria applied in this analysis; as described in Sec.~\ref{subsec:photon_identification}.
For 2011, the average number of $pp$ interactions
in the same bunch crossing is nine.
Events are required to have a reconstructed primary vertex (PV)
~\footnote{The leading primary vertex is defined as the vertex with the highest
$\sum|p_T|^2$ of tracks.}
consistent with the average beam-spot
position and with at least three associated tracks.

\subsection{Simulated events}
\label{subsec:simulated_data}
To study the characteristics of signal and background events,
MC samples are generated using {\tt PYTHIA} 6.4~\cite{pythia}, 
a LO  parton-shower MC generator, with the modified LO
 MRST2007~\cite{MRST2007}  PDFs.
The event generator parameters are set according to the ATLAS 
AMBT2 tune \cite{AMBT2}.
The ATLAS detector response is simulated using the {\tt GEANT4}
program~\cite{geant}.
In order to have a realistic description of the experimental
conditions under which the data are taken, pileup 
interactions are included in the simulation.
These samples are then reconstructed with the same algorithms
used for data. 
More details of the event generation and simulation 
infrastructure of the ATLAS experiment are provided in
Ref.~\cite{ATLAS_simulation}. 

For the study of systematic uncertainties and for comparisons with the final cross sections, 
events are generated with the {\tt HERWIG} 6.5~\cite{Herwig} model using the ATLAS
AUET2 tune~\cite{AUET2} and the same PDFs as used for the {\tt PYTHIA}
event generation.
{\tt HERWIG} and {\tt PYTHIA} use  different parton-shower and hadronization models.



Signal MC samples include hard-scattering photons from the LO
processes $q g \rightarrow q \gamma$ and $q \bar{q} \rightarrow g \gamma$,
and photons from QED radiation from quarks produced in QCD $2\to 2$ processes.

To study background processes, MC samples enriched in photons from 
meson decays with an $\ETg>100$~GeV  are used.
The samples are generated using all tree-level
2$\rightarrow$2 QCD processes, 
while events with photons originating from quarks were removed.


%% file: photon_reconstruction.tex
The reconstruction of photons in the ATLAS detector is described in
detail elsewhere ~\cite{Aad:2010sp, ATLAS_EP}.
The selection of photons is discussed in the following three sections:
kinematic preselection, isolation selection and shape
identification.

\subsection{Photon kinematic preselection}
\label{subsec:photon_preselection}
As already stated in Sec.~\ref{sec:Samples}, photon candidates are first 
required to have passed an 80~GeV trigger. From these, only those with
calibrated transverse energies 
above 100~GeV are retained for the subsequent analysis.
The calibration includes an in situ technique
based on the $Z$ boson mass peak~\cite{electronperf_2011}.
In order to benefit from the fine segmentation of the first
layer of the ECAL for identification of  
genuine photons,
the photon candidates are required to be within
the barrel  or the end-cap pseudorapidity regions.
After the selection, approximately
2.6$\times 10^6$ photon candidates remain in the data sample.


\subsection{Photon isolation selection}
\label{subsec:isolation}
Isolation is an important observable for prompt photon studies. 
The prompt photon signal is expected to be more isolated from hadronic
activity than the background.
Also, because of the mixture of hard-scattering and fragmentation
contributions in the prompt photon signal, it is important to have a
well modeled isolation variable that can be linked to the parton-level
isolation cut used in NLO QCD computations.  A
robust isolation prescription helps limit the non-perturbative fragmentation
contribution, which is poorly understood in theory, while retaining the
signal produced from direct processes.

This study uses the same definition of the cone isolation variable, $\ETiso$, as for  
the previous ATLAS measurement \cite{Aad:2011tw}. It is computed using
calorimeter cells from both the ECAL and HCAL, 
in a cone of radius $\Delta R=0.4$ around
the photon candidate.  
The contributions from the $5 \times 7$ second layer 
ECAL cells in the $\eta\times\phi$ space around the photon shower barycenter are not included in the
calculation. 
The expected small value of the leakage from the photon shower into the cone outside this small central region, evaluated as a
function of the $\ETg$ in simulated samples of single
photons, is then subtracted from the isolation variable.  
The contribution to the photon isolation energy from the underlying event 
and pileup 
is subtracted using the procedure proposed in Refs.~\cite{Cacciari:area, Cacciari:UE},
and implemented as described in Ref.~\cite{Aad:2010sp}.
After these corrections, the transverse isolation energy of simulated prompt photons 
is independent of $\ETg$.
A residual mild dependence on the amount of in-time pileup (from collisions of protons in the
same bunches as the  hard $pp$ scattering from which the photon originates) is observed for this isolation 
variable. 
This dependence can be traced back to the fact that $\ETiso$~is calculated from cells without
noise suppression whereas the pileup correction is computed from noise-suppressed topological clusters~\cite{topocluster}.
The pileup dependence of $\ETiso$ is well modeled in the simulation
and found to be robust against systematic uncertainties discussed later. 

In the following, all photon candidates having reconstructed isolation
energies $\ETiso\leq7$~GeV are considered ``isolated'',
while candidates with $\ETiso>7$~GeV are considered ``non-isolated''.
These definitions are applied to the data and to the MC calculations at 
both parton and particle level.
An ambient energy algorithm correction, which is used to correct for the activity of the underlying event, 
is also applied for the particle level MC isolation.
The isolation requirement $\ETiso \leq 7$~GeV is looser than that used in the previous analysis~\cite{Aad:2011tw} 
and is chosen in order to optimize the signal purity and the photon reconstruction efficiency at high $\ETg$.

\subsection{Photon shower-shape identification}
\label{subsec:photon_identification}

Shape variables computed from the lateral and longitudinal energy
profiles of the shower in the ECAL
are used to further discriminate the signal from the background. 
The selection criteria do not depend on the photon candidate's
$\ETg$, 
but vary as a function of the photon's reconstructed $\ETAg$ 
to take into account significant changes in the total
thickness of the upstream material and variations in the calorimeter
geometry or granularity.
The same ``tight'' selection criteria as used in the measurement of the isolated 
photon-pair production cross section~\cite{:2012je} are applied.
After the photon identification requirements, 1.3$\times 10^6$ (6.2$\times 10^5$) 
photon candidates remain in the 
barrel (end-cap) $\ETAg$ region. 
  The fraction of converted photons, i.e. photons that pair produce in the presence of 
material and are identified by their tracks are considered converted photons, is $32\%$ ($45\%$) in the barrel
(end-cap) $\ETAg$ region. 
Converted photons are identified by there tracks.
There are 
19 photon candidates with $\ETg$
between 800~GeV and 1~TeV.
The total number of events with more than one good photon candidate
is 1240.

%% file: purity.tex
The main background for prompt photons is due to hadronic jets containing $\pi^0$ mesons that 
carry most of the jet energy and that decay to photon pairs.
Such background photons are expected to be less isolated than prompt
photons due to activity from the other particles in the jet.
The isolation energy, $\ETiso$,  therefore provides a discrimination 
between prompt photons and photons from jets and meson decays.
To avoid relying on the simulation to accurately model the
energy flow inside jets and the fragmentation to $\pi^0$ mesons, a
data-driven technique is used for the reconstruction of 
the background isolation distribution. 

Among the shower-shape variables used in the photon
selection, some measured in the first ECAL
layer are fairly uncorrelated with the  $\ETiso$. 
A background-enhanced sample is provided by
requiring the photon candidates to fail the ``tight'' identification
criteria for at least one of four variables computed from the finely 
segmented first layer of the calorimeter,
and to satisfy all the other criteria \cite{Aad:2010sp}.
From now on, such photons are called
``non--tight candidates'', while the photon candidates satisfying the ``tight''
selection are called ``tight candidates''.

The residual background contamination in the 
tight candidates event sample is estimated using the 
``two-dimensional side bands'' method
\cite{Aad:2010sp}.
It is based on the definition of a ``tight--isolated'' signal
region $A$ and three background control regions $B$, $C$,
$D$: ``tight--non-isolated'', ``non-tight--isolated'' and ``non-tight--non-isolated'', respectively. 
The basic method assumes that the control regions have negligible signal
contamination and that the isolation energy distribution of background 
events is the same for tight and non-tight candidates. In that case the
signal yield in region $A$, $N_S^A$,
can be obtained from the number $N^{k}$ of events observed 
in data, in each of the four regions $k=A$, $B$, $C$, and $D$, as
\begin{equation}
N_{S}^A=N^{A}-N^{C}\frac{N^{B}}{N^{D}}.
\label{eq:purity_uncorrelated}
\end{equation}
The method can easily be extended to account for deviations from the previous
hypotheses, requiring only a limited knowledge of the signal and background 
properties. In that case, the equation to solve is 
\begin{equation}
N_{S}^{A} = N^{A}-R_{\rm BKG} \frac{(N^{B}-c_{B}N_S^{A})(N^{C}-c_{C}N_S^{A})}{(N^{D}-c_{D}N_S^{A})}; 
\label{eq:purity_correlated}
\end{equation}
where $c_{k}={N_{\rm S}^{k}}/{N_{\rm S}^{A}}$ are the fractions of signal 
events expected in  each of the three control regions, relative to the signal region $A$, and
$R_{\rm BKG} = {N_{\rm BKG}^{A}N_{\rm BKG}^{D}}/{N_{\rm BKG}^{B}N_{\rm BKG}^{C}}$ characterizes
the correlation between the isolation and identification variables in background
events ($R_{\rm BKG}=1$ when the correlations are negligible).

\begin{figure}[!htbp]
\begin{center}
\subfigure[]{
\includegraphics[scale=0.31, angle=0]{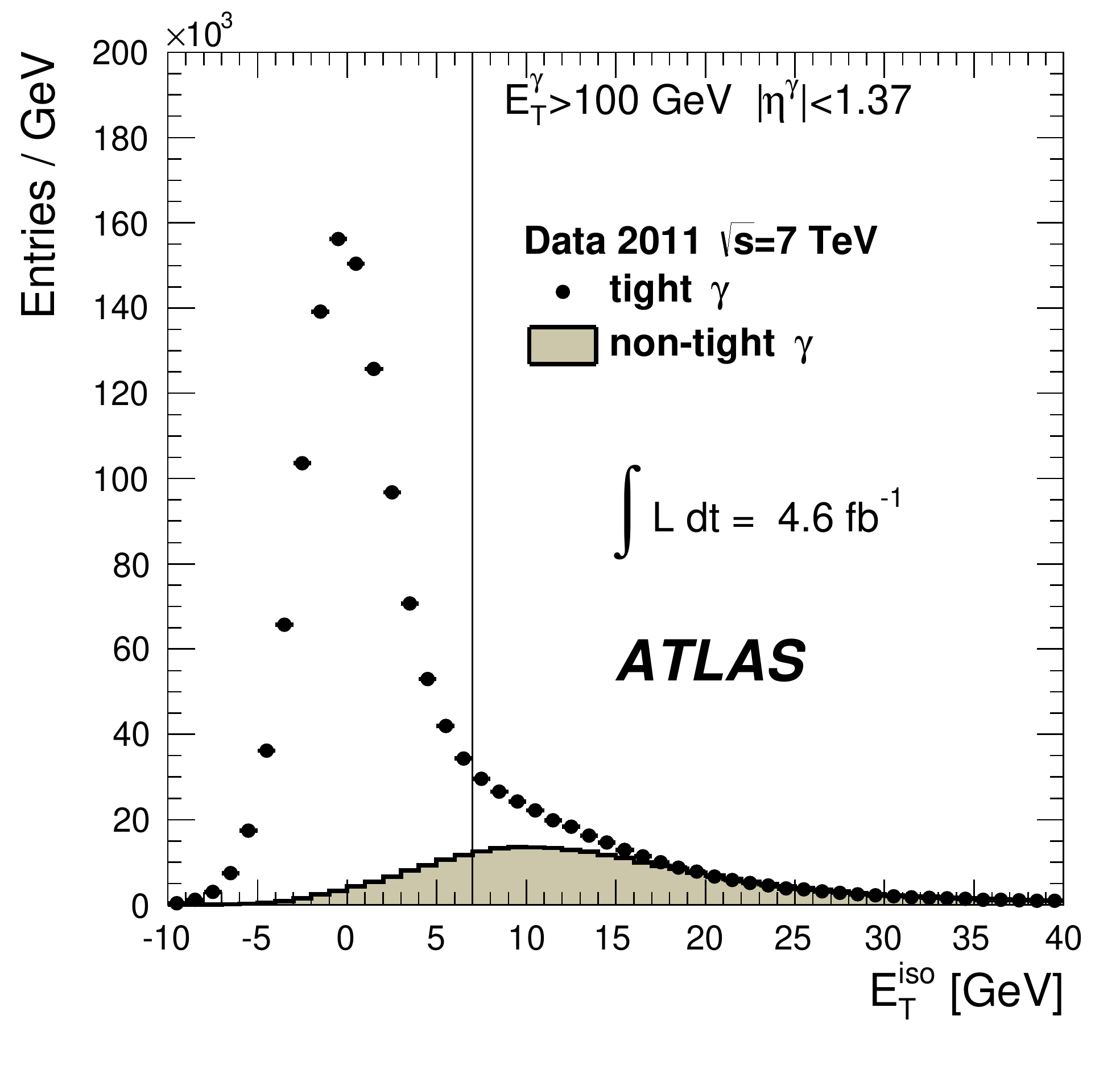}
}
\subfigure[]{
\includegraphics[scale=0.31, angle=0]{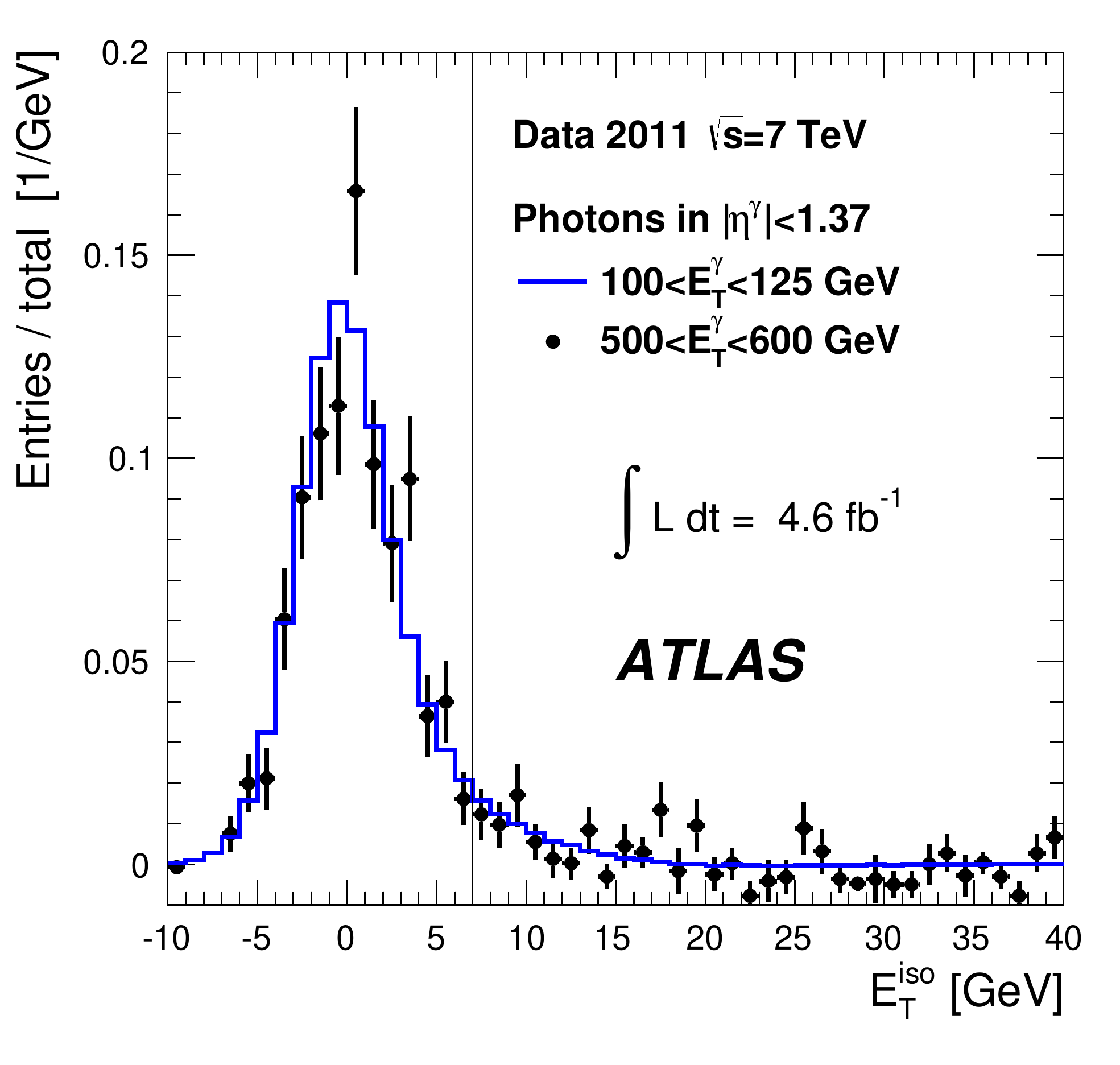}
}
\subfigure[]{
\includegraphics[scale=0.31, angle=0]{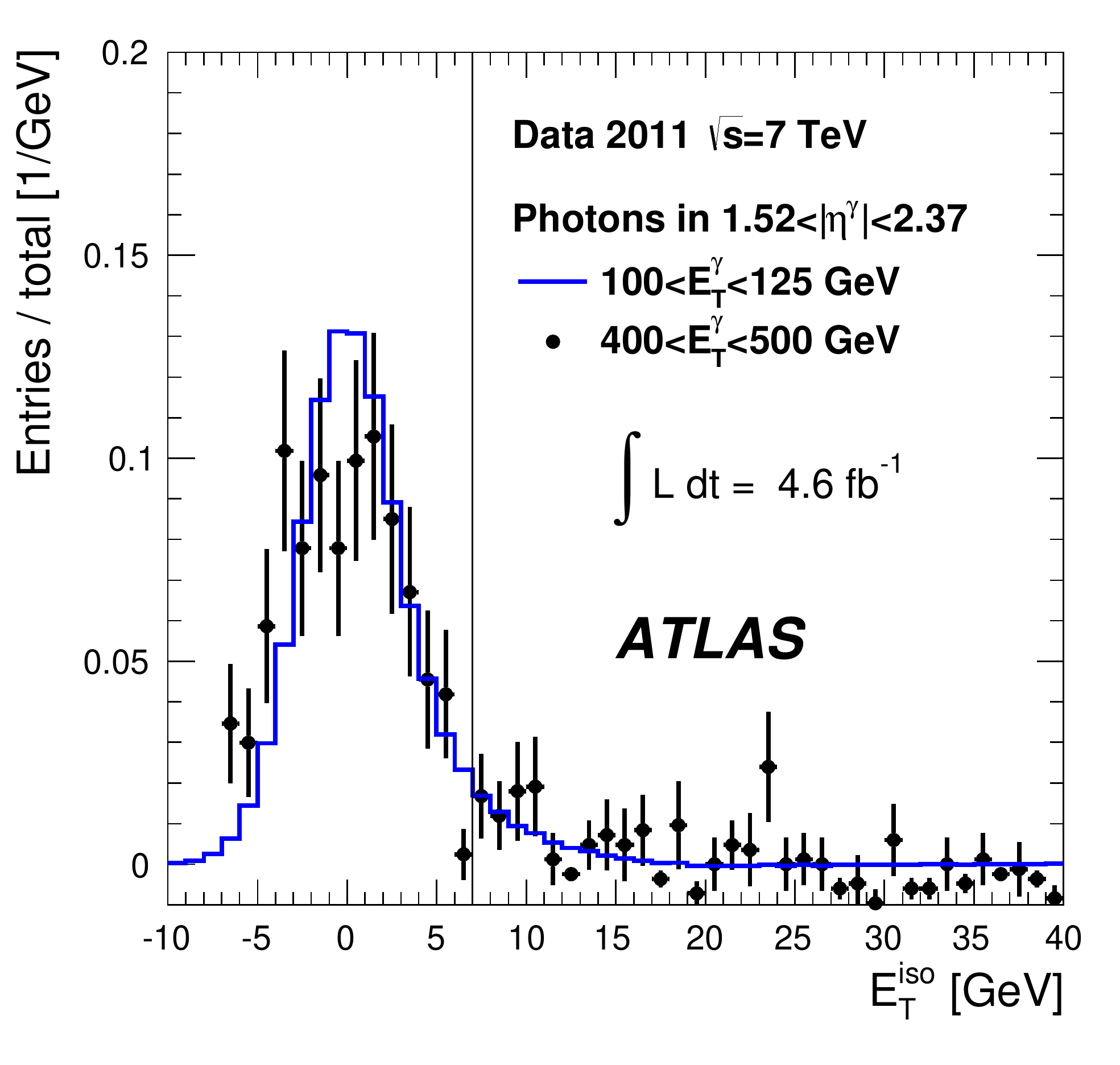}
}

\end{center}
\caption{(a): Distributions of tight 
photon transverse energy $\ETiso$~(dots) and 
  non-tight (shaded gray region) photon candidates in data, for photon transverse energy
  $\ETg>100$~GeV in the central $\ETAg$ region.
  The latter is normalized to the former for $\ETiso>15$~GeV.
  Distributions of tight $\ETiso$ photons 
  in the barrel (b) and end-cap (c) regions
  after subtracting the normalized non-tight distribution.
  For both (b) and (c)~a comparison of two representative $\ETg$ regions with different $\ETAg$ is shown.
  The vertical lines show the requirement of $\ETiso \leq 7$~GeV used to define the final cross sections.
  These distributions are normalized to one.
  }
 \label{fig:isoCentralPT}
\end{figure}

Figure~\ref{fig:isoCentralPT}(a) shows the distribution of $\ETiso$ for
tight and non-tight candidates. The latter is normalized to the former
in the background-dominated region $\ETiso>15$~GeV. 
The excess of tight candidates over normalized non-tight
candidates in the region $\ETiso<15$~GeV shows a clear peak
for signal prompt photons.
Fig.~\ref{fig:isoCentralPT}(b) and~\ref{fig:isoCentralPT}(c) show
the isolation profile of 
photon candidates after subtracting the distribution of non-tight 
candidates (with the same normalization as applied in Fig.~\ref{fig:isoCentralPT}(a)),
for different ranges of the photon candidate transverse energy in the two different~$\ETAg$ regions.
The distributions of these signal-enriched samples are
largely independent of the $\ETg$ range, according to the simulation.

In the following, Eq.~(\ref{eq:purity_correlated}) is used to estimate
the prompt photon yield in the selected sample, with $R_{\rm BKG}$ fixed to one as 
observed (within uncertainties) in simulated background events.
Results obtained neglecting signal leakage in the control regions, as in
Eq.~(\ref{eq:purity_uncorrelated}), or with $R_{\rm BKG}\ne 1$  are
used to evaluate systematic uncertainties.  In the end-cap region
there are too few events in the 500 to 600~GeV bin; therefore, 
the signal purity from the preceding bin is used instead.

\begin{figure}[!htbp]
\begin{center}
\includegraphics[scale=0.4, angle=0]{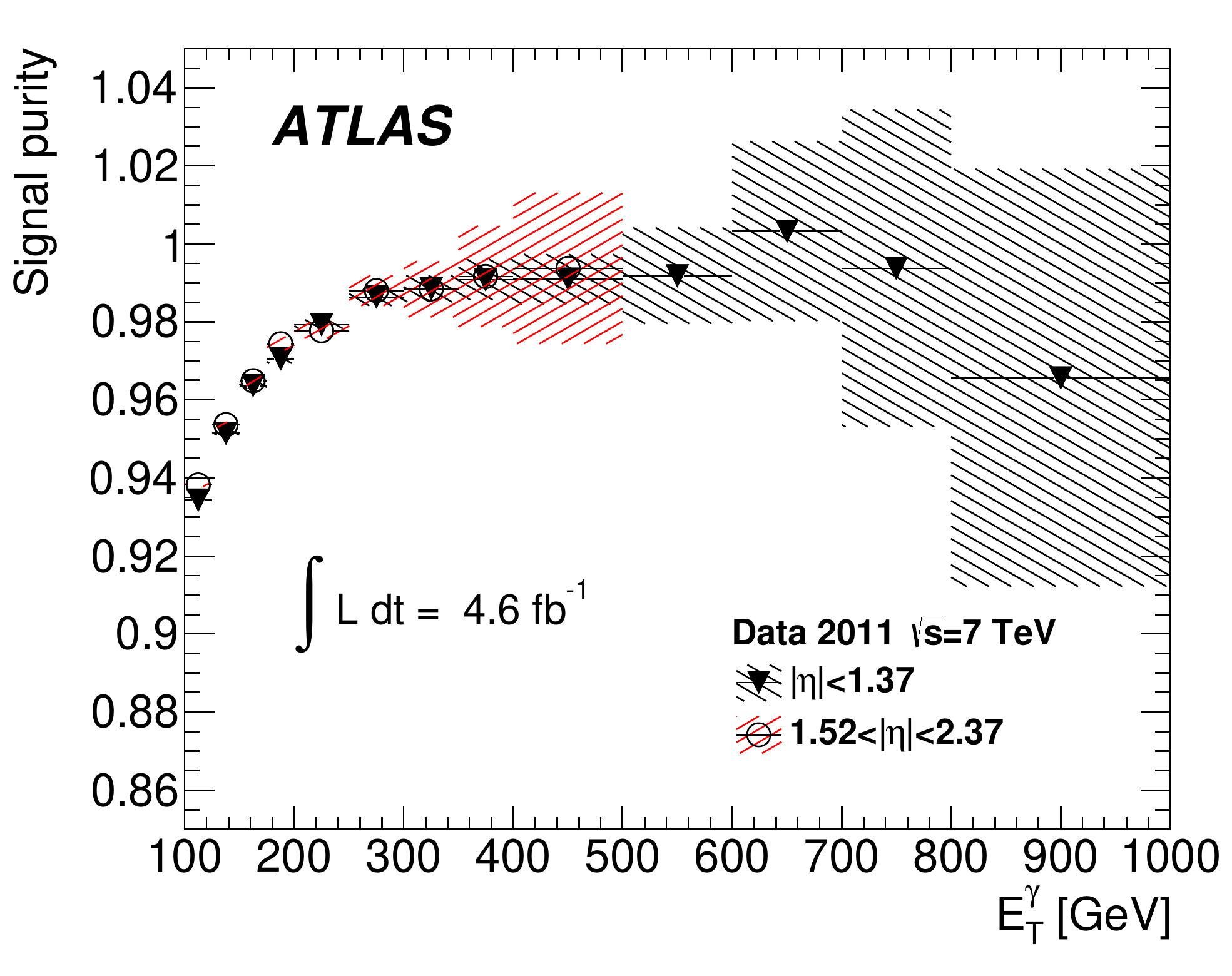}
\end{center}
\caption{The signal purity for the barrel
         and end-cap $\ETAg$ regions as a function of photon transverse energy~$\ETg$
         estimated from the data using the 
         two-dimensional side band approach shown in Eq.~(\ref{eq:purity_correlated}).
         The shaded
         bands indicate statistical uncertainties. 
         }
  \label{fig:purity1}
\end{figure}

The largest contribution to the impurity arises from background photons that come from the meson decays.
Figure~\ref{fig:purity1} shows the signal purity for prompt photons 
in region $A$ as a function of $\ETg$ for the barrel and end-cap regions.
The signal purity is estimated from the data using the two-dimensional side band 
approach shown in Eq.~(\ref{eq:purity_correlated}).
The shaded bands indicate statistical uncertainties.
The measured signal purity is 
larger than 93\% and increases with $\ETg$.
The purity has also been determined using Eq.~(\ref{eq:purity_correlated}) and the 
result agrees with the default method to within 3\% and has a similar $\ETg$-dependence.

%% file: unfolding.tex
A possible residual background could arise
from electrons that fake photons; primarily high-$p_{\rm T}$ isolated electrons
from $W$ or $Z$ boson decays that tend to be misidentified as converted photons.
The corresponding misidentification
probability is measured by studying the invariant mass spectrum of
$e^{\pm}\gamma$ combinations in the $Z$ boson mass range.
It was found that the background from prompt electrons is $\approx 0.5\%$ for $\ETg<400$~GeV  \cite{Aad:2011tw}.

To understand the background from electrons  for higher  $\ETg$,
a MC study was performed  using a sample of high-$p_{\rm T}$ electrons.
The current analysis neglects  contributions from high-$p_{\rm T}$ isolated electrons, since 
the MC study indicates that the rate of misidentified
photons with $\ETg$ above 400~GeV  originating from electrons   
is well below $0.5\%$.

%% file: efficiency.tex
The differential cross section for the production of isolated
prompt photons in a given phase-space bin $i$ is 
$N_i / (C_i(\gamma)\cdot  \Delta_i \cdot \int{\mathcal{L} dt})$, where
$N_i$ is the number of photons in a bin $i$ 
after the background subtraction, $C_i(\gamma)$ is a  correction factor,
$\Delta_i$ is the width of bin $i$ and  $\int{\mathcal{L} dt}$ 
is the integrated luminosity.
The correction factor, $C_i(\gamma)$, is evaluated from the bin-by-bin ratio
of the number of reconstructed prompt photons to the number of particle-level prompt photons in the signal simulation. 
The photon reconstruction efficiency in the MC simulation was tuned using data-driven techniques \cite{ATLAS-CONF-2012-123}.
The  correction factor $C_i(\gamma)$ accounts for 
acceptance and smearing effects, photon reconstruction efficiency and selection efficiency,
as well as the event selection efficiency.  
The various components of the correction are discussed.

\begin{itemize}

\item Acceptance and smearing correction, 
      is defined as the efficiency for a particle-level photon, 
      in the acceptance of the differential cross section, 
      to be reconstructed as a photon passing all the selection 
      criteria outlined in Sec.~\ref{sec:CrossSection}. 
      The largest contributing factor to this efficiency is the selection requirement $\ETiso\leq7$~GeV. 
      The shower-shape corrections for the MC simulation are determined from the comparison of
data with the simulation in the control
samples of photons selected in the same kinematic regions as used in this measurement.  
The average value of this efficiency in the barrel region was found to be $95\%$, 
while it is $87\%$ in the end-cap region.

\item Identification efficiency, which 
is defined as the efficiency for 
reconstructed  prompt photons after the isolation requirement
to pass the tight photon identification criteria 
described in Sec.~\ref{sec:Purity}. 
      This efficiency was estimated by using simulated signal events
      after correcting the simulated shower-shapes in the calorimeter to match those observed in data~\cite{Aad:2010sp}.
This efficiency in the barrel and 
end-cap region was found to be above $93\%$.

\item Trigger efficiency,  
      which is defined as  the efficiency for an event
      to be accepted by a photon trigger with an energy threshold of 80~GeV.
      The trigger efficiency is determined using a data-driven  technique based
      on high level triggers with low-$\ETg$  threshold, and it is estimated to be 
      close to $100\%$ for $\Etg>100$~GeV~\cite{trigger}.

\end{itemize}

In addition to the efficiencies quoted above, the correction factor 
also accounts for the bin-by-bin migration
due to the finite bin sizes. 
The MC simulations indicate that the RMS of the $\ETg$ resolution for photons
in the range $100<\ETg<600$~GeV is close to $3\%$ in the central region and
$4\%$ in the end-cap region.
The widths of the bins for the differential cross-section measurement are  chosen
to be substantially larger than the resolution in order to  minimize migration between neighboring bins. 

\begin{figure}[!htbp]
\begin{center}
\includegraphics[scale=0.4, angle=0]{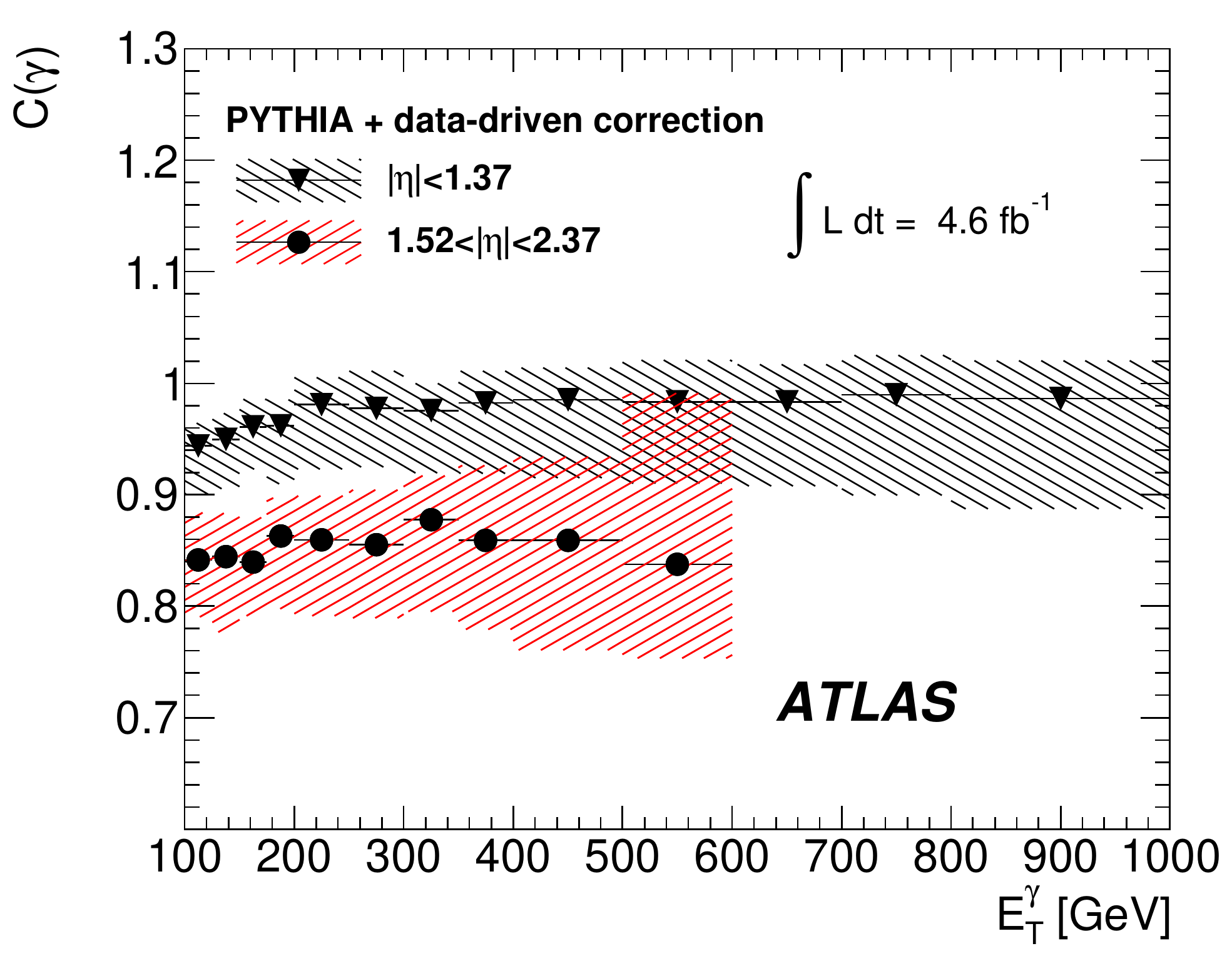}
\end{center}
\caption{The correction factor, $C_i(\gamma)$,  as a function of photon transverse energy $\ETg$ for the barrel and end-cap 
regions. The correction factor is evaluated from the bin-by-bin ratio, using
the {\tt PYTHIA} simulation,
of reconstructed prompt photons to particle-level prompt photons in the signal simulation.
The shaded
bands indicate statistical and systematic uncertainties 
discussed in Sec.~\protect{\ref{sec:uncertainties}}.
         }
  \label{fig:efficiency}
\end{figure}

The average value of the $C_i(\gamma)$ estimated using {\tt PYTHIA}   
is about 94\% in the barrel region and
86\% in the end-cap region. It 
increases with $\ETg$ by approximately 4\% in the range of $\ETg$ 
explored in this measurement.
This correction factor is shown in Fig.~\ref{fig:efficiency}, where the shaded bands represent the 
systematic and statistical uncertainties 
discussed in Sec.~\ref{sec:uncertainties}.

%% file: systematic_Uncertainties.tex
The systematic uncertainties on the measured differential cross sections 
are determined by 
repeating the analysis with some of the selection or analysis
procedures changed.
The systematic variations affect 
the $C_i(\gamma)$ correction factors and signal purity, thus 
an overall change in the cross section. 
The largest uncertainties are
described below.

\begin{itemize}

\item
A shift between the true and reconstructed isolation energy in the MC simulation was found to be
less than $700$~MeV  for  $\ETiso\simeq 7$~GeV.
This difference does not depend on the $\ETg$, and is similar
in \texttt{PYTHIA} and \texttt{HERWIG} signal and background MC samples.
This difference is also similar to that observed between the data and MC simulation.  
In the previous publication \cite{Aad:2011tw}, this difference estimated using electrons was found to be $500$~MeV.

MC samples with
an additional amount of material  in front of the calorimeter show a small effect on the isolation distribution.
For this MC, the shift between the true and reconstructed levels for the isolation is  
smaller than $700$~MeV. The correction factors $C_i(\gamma)$ calculated using such MC showed a negligible effect on the cross section. 
 
The systematic uncertainty on the cross section due to the isolation cut 
was evaluated by changing the requirement by $\pm 700$~MeV in the simulation and
recalculating the correction factors $C_i(\gamma)$.
This systematic variation leads to a typical uncertainty of less than $1\%$ for all $\ETg$
explored in this measurement.

\item
The uncertainty on the cross section due to insufficient knowledge of
the photon identification efficiency is
estimated by using different techniques for the
photon identification as described in Ref.~\cite{ATLAS-CONF-2012-123}.
Such uncertainties also account for the amount of material upstream of the calorimeter.
An effect of 2\% or less is observed for all $\ETg$
explored in this measurement.

\item
The uncertainty due to the photon energy measurement   
is calculated by varying the photon
energy scale within the expected uncertainty in the MC simulation. 
This uncertainty mostly affects the $C_i(\gamma)$ correction factor.
The effect of such a variation leads to an uncertainty between 2\% at low $\ETg$ and 6\% at large $\Etg$.  

\item
The systematic uncertainty on the cross section due to
the photon energy resolution is calculated
by smearing the central value and then varying the reconstructed energy in the MC simulations as described in Ref.~\cite{Aad:2010sp} and then
recomputing the $C_i(\gamma)$ factor.
This uncertainty is typically 2\% for all $\ETg$
explored in this measurement.

\item
The stability on the $C_i(\gamma)$ factors due to the choice of MC generator is computed by
considering \texttt{HERWIG} for the bin-by-bin correction instead of \texttt{PYTHIA}. 
The stability affects photon reconstruction and identification.
It also probes the uncertainty on the signal reconstruction due to an alternative fragmentation mechanism.
The uncertainty on the cross section due to this contribution ranges from $2\%$ at low $\Etg$ to $4\%$ at $\Etg>800$~GeV.

\item
The uncertainty on the background subtraction is estimated using alternative
background subtraction techniques discussed in Sec.~\ref{sec:Purity}.
Equation~(\ref{eq:purity_correlated}) is modified to either neglect signal leakage or
include a modified $R_{\rm BKG}$.
The background is subtracted by either  
neglecting correlations between the signal and background regions, or using
the central values of the correlations estimated from simulated background
events. 
The uncertainty on the cross section due to the background subtraction technique
varies between 2\% and 3\% for all $\ETg$ explored in this measurement.

\item
The uncertainty arising from the definition of the background control regions 
is estimated by repeating the measurement using an  alternative
definition of the non-isolated region.
The isolation requirement was increased from 7~GeV to 10~GeV.
Such a redefinition affects both the signal purity and the $C_i(\gamma)$ factors. 
An effect of 1\% or less for all $\ETg$ 
explored in this measurement is observed, which is compatible with the statistical uncertainty.


\item
The systematic uncertainty on the fraction of photons from fragmentation
was estimated using the PYTHIA signal sample with 
$50\%$ fewer photons from fragmentation.
Alternatively, weights of events with photons from fragmentation were scaled by a factor of two. 
The effect from such changes on the final cross sections 
is compatible with the statistical uncertainty ($<0.5\%$).

\item
The relative systematic uncertainty on the cross section due to the uncertainty of 
the luminosity measurement is $1.8\%$. It is fully correlated among all 
$\ET$ and $\eta$ bins of the differential cross sections. 
\end{itemize}

The sources of systematic uncertainty
are assumed uncorrelated and thus the total
systematic uncertainty
is estimated by summing in quadrature all the contributions.
The final systematic uncertainty on the differential and total 
cross sections in the barrel (end-cap) region is below  $6 (7)\%$.

As a cross-check, the  measurement is repeated using an alternative
definition of the photon transverse isolation energy, based on
three-dimensional topological clusters~\cite{topocluster} of energy deposits in the
calorimeters, affecting mostly the photon reconstruction efficiency. 
The same calorimeter cells are used for both the
calculation of the photon isolation and for the subtraction of
the contribution from the underlying event and pileup, thus
providing a quantity that is less dependent on the amount of pileup.
A difference smaller than 3\% is found between the alternative
and the nominal results.
In addition,  in order to verify the reliability of the pileup removal technique,
differential cross sections  were calculated separately for low pileup and high-pileup runs.
These two cross sections were found to be consistent.

%% file: theory.tex

The expected prompt-photon production cross section
was estimated using the {\tt JETPHOX}~1.3 Monte Carlo
program~\cite{jetphox1,PhysRevD.73.094007}, which implements a full
NLO QCD calculation of both the direct and fragmentation contributions
to the total cross section.  The parton-level isolation, defined as the
total $E_{\mathrm T}$ from the partons produced with the
photon inside a cone of radius $\Delta R=0.4$
in $\eta\times\phi$ around the photon direction,
is required to be smaller than 7 GeV.
The fragmentation contribution in the {\tt JETPHOX} calculation decreases
with increasing $\Etg$ and becomes negligible for $\Etg>500$~GeV.  
Further details of the {\tt JETPHOX} calculation can be found in Ref.~\cite{Blair:1296265}.
The calculation uses the NLO photon fragmentation function of BFG set II~\cite{Bourhis:1998}.
The CT10~\cite{Lai:2010vv} and MSTW2008NLO~\cite{Martin:2009iq} PDFs for the
proton are provided by the
{\tt LHAPDF} package~\cite{Giele:2002hx}. 
The nominal renormalization ($\mu_R$), factorization ($\mu_F$) and
fragmentation ($\mu_f$) scales were set to the photon transverse
energy ($\mu_R = \mu_F = \mu_f = \ET^\gamma$). 
Systematic uncertainties on the QCD cross sections are estimated
and listed below.
\begin{itemize}
\item The scale uncertainty is evaluated by varying the three
scales following the constraints and are added in quadrature~\cite{Blair:1296265}:
  \begin{itemize}
  \item $\mu_R = \mu_F = \mu_f \in[\ET^\gamma/2,  2\ET^\gamma]$;
  \item $\mu_R \in[\ET^\gamma/2, 2 \ET^\gamma], \mu_F = \mu_f = \ET^\gamma$;
  \item $\mu_F \in[\ET^\gamma/2, 2 \ET^\gamma], \mu_R = \mu_f = \ET^\gamma$;
  \item $\mu_f \in[\ET^\gamma/2, 2 \ET^\gamma], \mu_R = \mu_F = \ET^\gamma$;
  \end{itemize} 
This leads to a change of between 12\% and 20\% in the predicted cross section.  

\item 
The uncertainty on the differential cross section due to insufficient knowledge of the PDFs
was obtained by repeating the {\tt JETPHOX} calculation for
52 eigenvector sets of the CT10 PDF and applying a scaling factor in order to  obtain the
uncertainty for the 68\% confidence-level (CL) interval~\cite{Blair:1296265}. 
The corresponding uncertainty on the cross section increases with $\ETg$ 
and varies between a 5\% at $\ETg\simeq 100$~GeV and 15\% at $\ETg\simeq 900$~GeV. 

\item
The effect of the uncertainty on the
value of the strong coupling constant, $\alpha_s$, is evaluated
following the recommendation in Ref.~\cite{Lai:2010vv}. 
This was done using different CT10 PDF sets with $\alpha_s$ values varied by $\pm 0.002$ around
the central value,  $\alpha_s=0.118$. 
Then, a scaling factor was applied in order to  obtain the
uncertainty for the 68\% CL interval.  The average $\alpha_s$ uncertainty on the cross section 
is $4.5\%$,
with a small dependence on $\ETg$.  
\end{itemize}

In the following, the total uncertainty includes the three sources 
above added in quadrature. In addition the uncertainty arising from the scale variations, 
which is the largest of these three contributions, will be shown separately.

In order to perform a proper comparison with the {\tt JETPHOX} calculation, the effects of hadronization,
pileup and the underlying event have to be understood because the isolation energy is directly sensitive to
these effects. 
The ambient-energy-density correction used for the $\ETiso$ reconstruction 
reduces the effects from the underlying event and pileup, but this effect may not be completely taken into account.
Using {\tt PYTHIA} and {\tt HERWIG} with different tunes, the combined effect 
from hadronization and the underlying event is estimated to be about $\pm 1\%$. 
This correction is small compared to the full uncertainty from other sources and is not included
in the total theoretical uncertainty. 

The measured cross sections are also compared to those from the
LO parton-shower generators, {\tt PYTHIA}  and {\tt HERWIG}.  These models are described   
in Sec.~\ref{subsec:simulated_data}.
Both simulate the fragmentation components through the emission of
photons in the parton shower.

%% file: theorycomparison.tex
\begin{figure}[h]
  \centering
  \includegraphics[width=0.9\columnwidth]{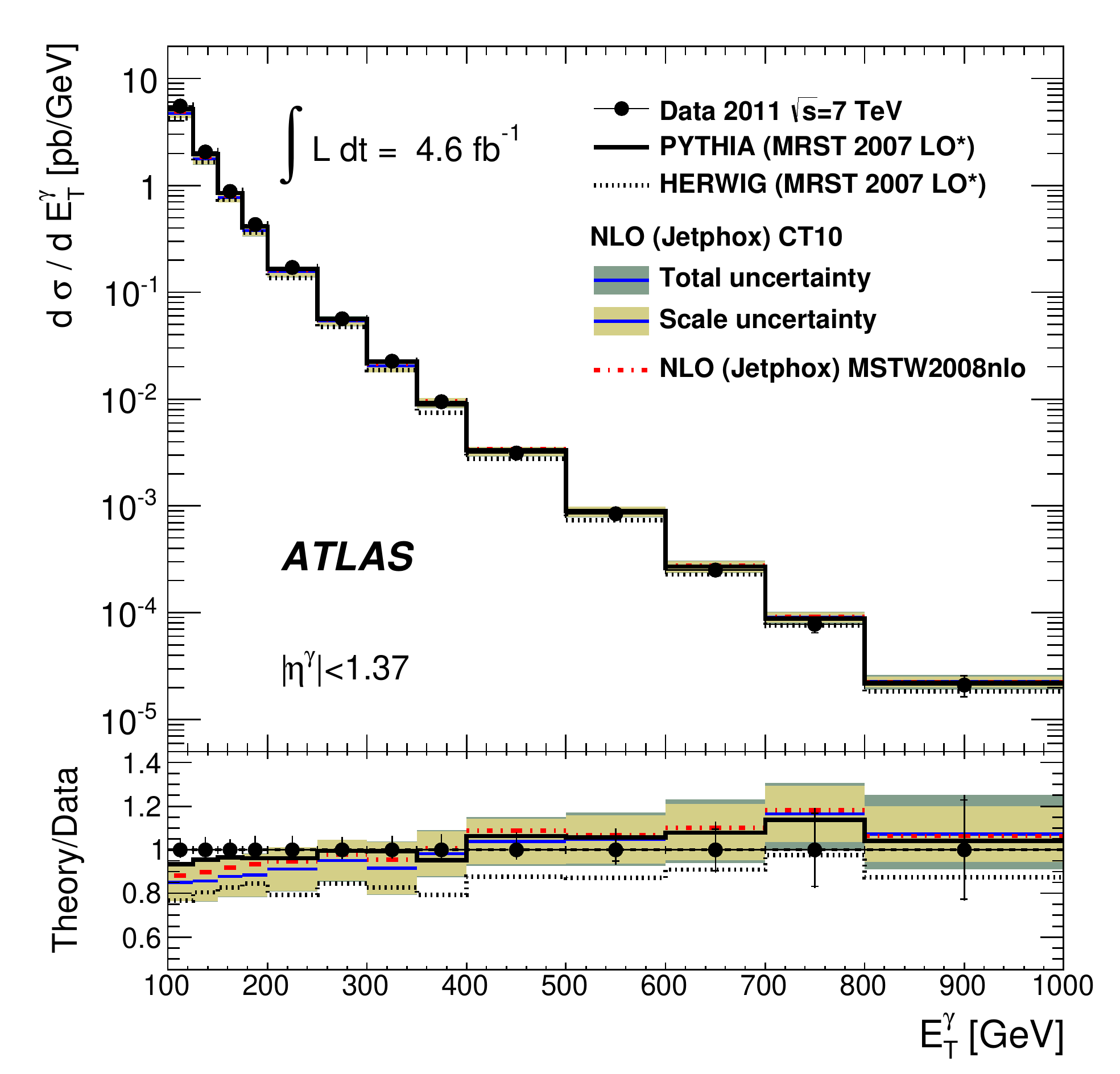}%
  \caption{Measured (dots with error bars) and expected inclusive 
  prompt photon cross section as a function of the photon transverse energy, $E_T^{\gamma}$,
  in the barrel $\ETAg$ region. 
  The inner error bars on the data points show statistical uncertainties, 
  while the full error bars show statistical and systematic uncertainties added in quadrature.
  The NLO theory prediction is shown with the shaded bands that indicate 
  the scale uncertainty (the inner yellow band) and the total uncertainty (the outer green band), which also includes the PDF and $\alpha_s$ uncertainties. 
  The  LO parton-shower MC generators
  are shown as lines.   The bottom panel shows the corresponding theory/data ratio, in which the data points
  are centered at one. 
  }
  \label{fig:xsec_measuredpt}
\end{figure}
\begin{figure}[!htbp]
  \centering
  \includegraphics[width=0.9\columnwidth]{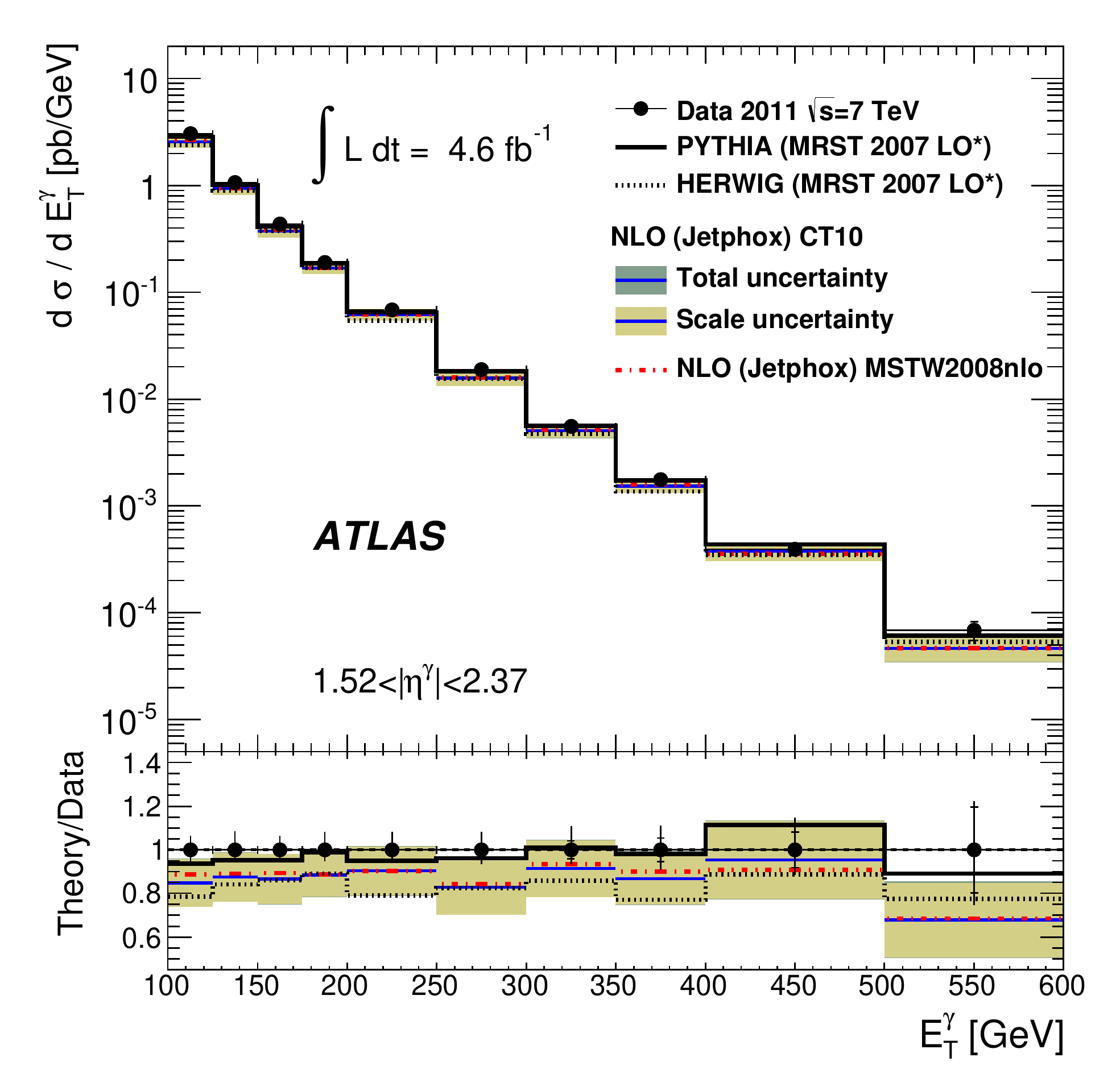}
  \caption{Measured (dots with error bars) and expected inclusive prompt photon 
    cross section
    in the end-cap region.
   The inner error bars on the data points 
   show statistical uncertainties, while the full error bars show statistical and systematic uncertainties added in quadrature.
  The NLO theory prediction is shown with the shaded bands that indicate
  the scale uncertainty (the inner yellow band) and the total uncertainty (the outer green band), which also includes the PDF and $\alpha_s$ uncertainties.  
The  LO parton-shower MC generators
  are shown as lines.   
}   
  \label{fig:xsec_measuredptF}
\end{figure}
\begin{figure}[!htbp]
  \centering
  \includegraphics[width=0.9\columnwidth]{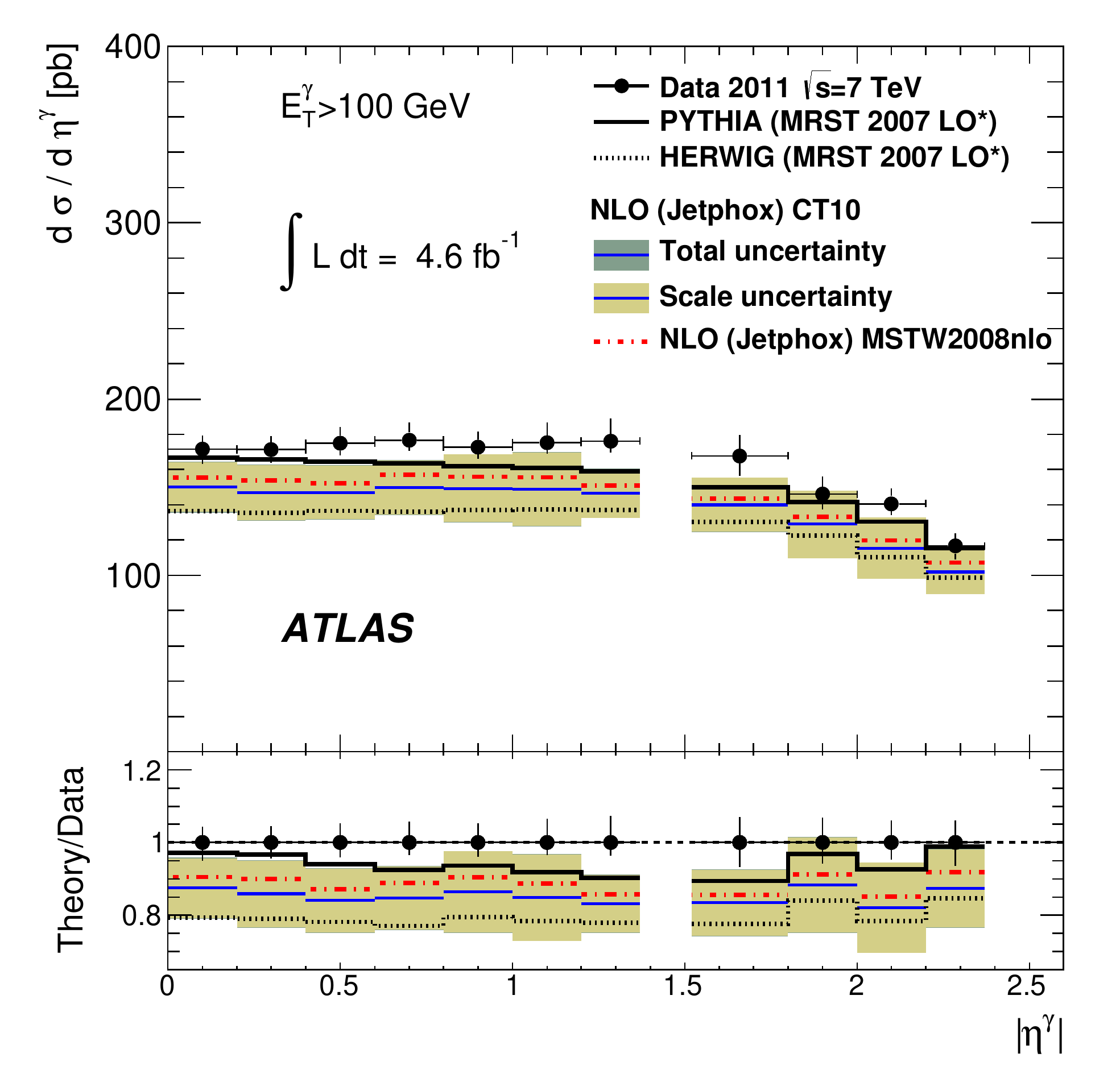}%
  \caption{Measured and expected inclusive prompt photon
    cross section as a function of $|\ETAg|$, for photons with 
    transverse energies above
    $100$~GeV excluding $1.37<|\ETAg|<1.52$.
    The data points 
    show full error bars that contain statistical, systematic, and luminosity
    uncertainties added in quadrature, and are negligible.
  The NLO theory prediction is shown with the shaded bands that indicate
  the scale uncertainty (the inner yellow band) and the total uncertainty (the outer green band), which also includes the PDF and $\alpha_s$ uncertainties.
The LO parton-shower MC generators
  are shown as lines.   
}   
\label{fig:xsec_measured_eta}
\end{figure}

The differential cross section for the production of isolated
prompt photons is obtained from the number of signal
events as discussed in Sec.~\ref{sec:efficiency}.
The measured $\ETg$-differential cross sections together with the
theoretical predictions are shown in
Figs.~\ref{fig:xsec_measuredpt} and \ref{fig:xsec_measuredptF} for
the barrel and end-cap $\ETAg$ regions, respectively.
Tables~\ref{tab_final_gamma_pt} and \ref{tab_final_gamma_pt_F}
list the values of the differential cross sections shown
in these figures.
Figure~\ref{fig:xsec_measured_eta} and Table~\ref{tab_final_gamma_eta} show the cross
section as a function of $\ETAg$ for $\ETg>100$~GeV. 
The full error bars on the data points represent the combination of
statistical and systematic uncertainties.
The inner error bars
show statistical uncertainties. 
The shaded bands on the NLO predictions show the theoretical uncertainties as discussed in Sec.~\ref{sec:Theory}.
The theoretical uncertainties due to the choice of factorization and renormalization
scales as well as the fragmentation scale are shown as an inner band.

The NLO calculations agree with the data up to the highest $\ETg$ considered. 
The data are somewhat higher than the central NLO calculation for low $\ETg$, 
but agree within the theoretical uncertainty of the
NLO calculation. 
This trend is also visible throughout $\ETAg$
as it is dominated by the low $\ETg$ range of the measurement. 
At low $E_T^{\gamma}$, the observed difference between the NLO
predictions based CT10 PDF and MSTW2008NLO PDF are larger than the PDF
uncertainty estimated using CT10. The difference between CT10  and
MSTW2008NLO predictions is smaller than the CT10 PDF uncertainty for
$E_T^{\gamma}>600$~GeV.

The predictions of the LO parton-shower MC generators, {\tt PYTHIA} and {\tt HERWIG}, are also
shown in Figs.~\ref{fig:xsec_measuredpt}
-\ref{fig:xsec_measured_eta}.
The {\tt PYTHIA} model describes the data fairly well, while  {\tt HERWIG} 
falls below the data by ~10\%-20\%.
The shapes of the cross
sections are well described by both models.

\begin{figure}[h]
\centering
\includegraphics[width=0.9\columnwidth]{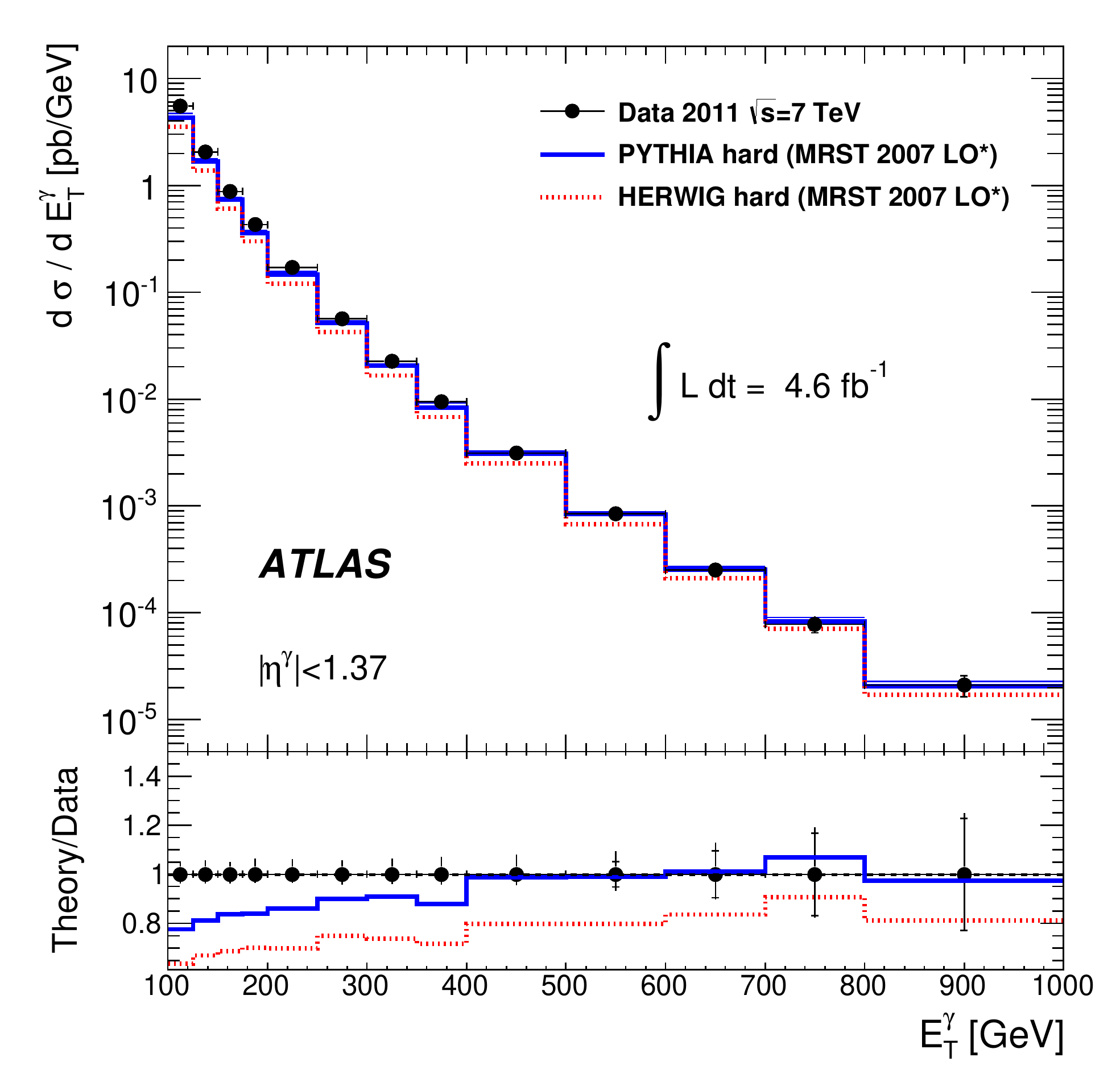}
\caption{
Same data as in Fig.~\ref{fig:xsec_measuredpt}, but the comparison is made with 
MC predictions that include only direct photons from the hard processes.}
\label{fig:final_gamma_hardPT}
\end{figure}

The data are also compared to MC predictions that include only direct photons 
from $q g \rightarrow q \gamma$ and $q \bar{q} \rightarrow g \gamma$ processes
calculated at LO QCD.
Figure~\ref{fig:final_gamma_hardPT} shows that these MC generators 
predict a cross section at low $\ETg$ that is 20\% lower than the data
which includes all the higher-order fragmentation processes. 
This difference is reduced at high $\ETg$, where the contribution from 
photons originating from fragmentation becomes small. This shows that the higher order fragmentation processes
contribute significatly to the shape of the predicted $\ETg$ cross
section.

The total inclusive cross section of direct photons calculated in the
kinematic region
$\ETg>100$~GeV, $|\ETAg|<1.37$ and 
$\ETiso\leq7$~GeV is
$$
\sigma(\gamma + X) = 236 \pm  2~(\mathrm{stat})^{+13}_{-9}
(\mathrm{syst}) \pm  4~(\mathrm{lumi})~\mathrm{pb}. 
$$
{\tt PYTHIA} predicts that this cross section is 224~pb while {\tt HERWIG} predicts 187~pb.
The cross section was calculated from the total number of signal events in the given kinematic region.
The NLO calculations with the CT10 and MSTW2008NLO PDFs 
predict $203\pm 25$~(theory)~pb and $212\pm 24$~(theory)~pb, respectively, where
the theory uncertainty
is symmetrized and includes the scale, PDF and $\alpha_s$ uncertainties.

The total inclusive cross section for direct photons within the kinematic range 
$\ETg>100$~GeV, $1.52\leq|\ETAg|<2.37$ and $\ETiso\leq7$~GeV is
$$
\sigma(\gamma + X) = 123 \pm  1~(\mathrm{stat})^{+9}_{-7}
(\mathrm{syst}) \pm  2~(\mathrm{lumi})~\mathrm{pb},
$$ 
which can be compared to predictions of 118~pb ({\tt PYTHIA}) and 99~pb ({\tt HERWIG}).
The NLO calculations based on CT10 and MSTW2008NLO PDFs 
predict $105\pm 15$~(theory)~pb and $109\pm 15$~(theory)~pb, respectively.

%% file: conclusion.tex
A measurement of the differential cross sections for the inclusive
production of isolated prompt photons in $pp$
collisions at a center-of-mass energy of $\sqrt{s} = 7$~TeV  was 
presented using 4.6~\ifb\ of collision data collected with the ATLAS
detector at the LHC.
The cross sections were measured as a function of photon transverse 
energy $\ETg$  and pseudorapidity $\ETAg$. 
The  $\ETg$  kinematic 
range of this measurement spans  from $100$~GeV to 1~TeV, 
thus significantly      
extending the measured kinematic range previously published~\cite{Aad:2011tw} by ATLAS.
The measured differential cross section falls by more than five orders of magnitude in this
kinematic range.

Both {\tt PYTHIA} and {\tt HERWIG} describe the shapes of the differential cross sections.
The  {\tt HERWIG} generator predicts a smaller cross section compared to {\tt PYTHIA} and the data. 
The MC studies presented in this paper indicate that 
fragmentation contributions are needed for a good description of the data.

The data agree with the NLO predictions based on the CT10 and MSTW2008 PDF up to the highest 
measured $\ETg \simeq 1$ TeV. In this kinematic regime,  
the theoretical uncertainties due to the PDFs of the proton become significant.  
Thus the presented cross sections have the potential to provide additional constraints on the 
proton PDFs.

%% file: tables.tex
\begin{table}[h]
\centering
   {\small
   \input{tab1}   
}
\caption{
Measured inclusive prompt photon production
cross section in the pseudorapidity range  $|\ETAg|<1.37$ 
as a function of $\ETg$ with statistical and systematic uncertainties.
}
\label{tab_final_gamma_pt}
\end{table}

\begin{table}[h]
\centering
   {\small
   \input{tab2}
}
\caption{
Measured inclusive prompt photon production
cross section 
in the pseudorapidity range $1.52\leq |\ETAg| < 2.37$ 
as a function of $\ETg$ with statistical and systematic uncertainties. 
}
\label{tab_final_gamma_pt_F}
\end{table}

\begin{table}[h]
\centering
   {\small
   \input{tab3}   
}
\caption{
Measured inclusive prompt photon production
cross section for $\ETg>100$~GeV as a function of $|\ETAg|$ 
with statistical and systematic uncertainties. 
}
\label{tab_final_gamma_eta}
\end{table}

%% file: tab1.tex
\renewcommand{\arraystretch}{1.3}
\begin{tabular}{|c|cccc|}
\hline 
$\ETg$ bin [GeV] &  $d\sigma / d \ETg$ &  $\pm$  (stat.) & $\pm$ (syst.) & [pb/GeV]  \\ \hline 
 $100-125$  & 5.55 & $ \pm 0.02$ & $^{+0.30}_{-0.21}$ &    \\ \hline
 $125-150$  & 2.06 & $ \pm 0.01$ & $^{+0.12}_{-0.08}$ &    \\ \hline
 $150-175$  & 8.82 & $ \pm 0.07$ & $^{+0.45}_{-0.32}$ & $\times 10^{-01}$  \\ \hline
 $175-200$  & 4.28 & $ \pm 0.05$ & $^{+0.27}_{-0.15}$ & $\times 10^{-01}$  \\ \hline
 $200-250$  & 1.71 & $ \pm 0.01$ & $^{+0.11}_{-0.06}$ & $\times 10^{-01}$  \\ \hline
 $250-300$  & 5.65 & $ \pm 0.07$ & $^{+0.33}_{-0.24}$ & $\times 10^{-02}$  \\ \hline
 $300-350$  & 2.25 & $ \pm 0.04$ & $^{+0.14}_{-0.09}$ & $\times 10^{-02}$  \\ \hline
 $350-400$  & 9.43 & $ \pm 0.21$ & $^{+0.67}_{-0.40}$ & $\times 10^{-03}$  \\ \hline
 $400-500$  & 3.12 & $ \pm 0.08$ & $^{+0.26}_{-0.14}$ & $\times 10^{-03}$  \\ \hline
 $500-600$  & 8.44 & $ \pm 0.44$ & $^{+0.82}_{-0.58}$ & $\times 10^{-04}$  \\ \hline
 $600-700$  & 2.50 & $ \pm 0.24$ & $^{+0.32}_{-0.26}$ & $\times 10^{-04}$  \\ \hline
 $700-800$  & 7.77 & $ \pm 1.30$ & $^{+1.49}_{-1.36}$ & $\times 10^{-05}$  \\ \hline
 $800-1000$  & 2.11 & $ \pm 0.48$ & $^{+0.53}_{-0.49}$ & $\times 10^{-05}$  \\ \hline
\end{tabular} 

%% file: tab2.tex
\renewcommand{\arraystretch}{1.3}
\begin{tabular}{|c|cccc|}
\hline 
$\ETg$ bin [GeV] &  $d\sigma / d \ETg$ &  $\pm$  (stat.) & $\pm$ (syst.) & [pb/GeV]  \\ \hline 
 $100-125$  & 3.03 & $ \pm 0.01$ & $^{+0.19}_{-0.19}$ &    \\ \hline
 $125-150$  & 1.06 & $ \pm 0.01$ & $^{+0.09}_{-0.06}$ &    \\ \hline
 $150-175$  & 4.34 & $ \pm 0.05$ & $^{+0.27}_{-0.24}$ & $\times 10^{-01}$  \\ \hline
 $175-200$  & 1.90 & $ \pm 0.03$ & $^{+0.16}_{-0.10}$ & $\times 10^{-01}$  \\ \hline
 $200-250$  & 6.84 & $ \pm 0.08$ & $^{+0.57}_{-0.37}$ & $\times 10^{-02}$  \\ \hline
 $250-300$  & 1.89 & $ \pm 0.04$ & $^{+0.16}_{-0.12}$ & $\times 10^{-02}$  \\ \hline
 $300-350$  & 5.52 & $ \pm 0.22$ & $^{+0.60}_{-0.36}$ & $\times 10^{-03}$  \\ \hline
 $350-400$  & 1.76 & $ \pm 0.10$ & $^{+0.19}_{-0.16}$ & $\times 10^{-03}$  \\ \hline
 $400-500$  & 3.93 & $ \pm 0.32$ & $^{+0.58}_{-0.46}$ & $\times 10^{-04}$  \\ \hline
 $500-600$  & 6.83 & $ \pm 1.35$ & $^{+1.53}_{-1.74}$ & $\times 10^{-05}$  \\ \hline
\end{tabular} 

%% file: tab3.tex
\renewcommand{\arraystretch}{1.3}
\begin{tabular}{|c|cccc|}
\hline 
$|\ETAg|$ bin &  $d\sigma / d |\ETAg|$ &  $\pm$  (stat.) & $\pm$ (syst.) & [pb]  \\ \hline 
 $0.0-0.2$  & 1.72 & $ \pm 0.01$ & $^{+0.08}_{-0.09}$ & $\times 10^{+02}$  \\ \hline
 $0.2-0.4$  & 1.71 & $ \pm 0.01$ & $^{+0.08}_{-0.08}$ & $\times 10^{+02}$  \\ \hline
 $0.4-0.6$  & 1.75 & $ \pm 0.01$ & $^{+0.09}_{-0.07}$ & $\times 10^{+02}$  \\ \hline
 $0.6-0.8$  & 1.77 & $ \pm 0.01$ & $^{+0.10}_{-0.06}$ & $\times 10^{+02}$  \\ \hline
 $0.8-1.0$  & 1.73 & $ \pm 0.01$ & $^{+0.09}_{-0.07}$ & $\times 10^{+02}$  \\ \hline
 $1.0-1.2$  & 1.75 & $ \pm 0.01$ & $^{+0.11}_{-0.06}$ & $\times 10^{+02}$  \\ \hline
 $1.2-1.37$  & 1.76 & $ \pm 0.01$ & $^{+0.13}_{-0.06}$ & $\times 10^{+02}$  \\ \hline
 $1.52-1.8$  & 1.68 & $ \pm 0.01$ & $^{+0.12}_{-0.11}$ & $\times 10^{+02}$  \\ \hline
 $1.8-2.0$  & 1.46 & $ \pm 0.01$ & $^{+0.10}_{-0.08}$ & $\times 10^{+02}$  \\ \hline
 $2.0-2.2$  & 1.41 & $ \pm 0.01$ & $^{+0.09}_{-0.07}$ & $\times 10^{+02}$  \\ \hline
 $2.2-2.37$  & 1.17 & $ \pm 0.01$ & $^{+0.07}_{-0.07}$ & $\times 10^{+02}$  \\ \hline
\end{tabular} 

%% file: atlas_authlist.tex
\begin{flushleft}
{\Large The ATLAS Collaboration}

\bigskip

G.~Aad$^{\rm 48}$,
T.~Abajyan$^{\rm 21}$,
B.~Abbott$^{\rm 112}$,
J.~Abdallah$^{\rm 12}$,
S.~Abdel~Khalek$^{\rm 116}$,
A.A.~Abdelalim$^{\rm 49}$,
O.~Abdinov$^{\rm 11}$,
R.~Aben$^{\rm 106}$,
B.~Abi$^{\rm 113}$,
M.~Abolins$^{\rm 89}$
% ...
\bigskip
\\
% ...
$^{11}$ Institute of Physics, Azerbaijan Academy of Sciences, Baku, Azerbaijan\\
$^{12}$ Institut de F{\'\i}sica d'Altes Energies and Departament de F{\'\i}sica de la Universitat Aut{\`o}noma de Barcelona, Barcelona, Spain\\
% ...
$^{21}$ Physikalisches Institut, University of Bonn, Bonn, Germany\\
% ...
$^{48}$ Fakult{\"a}t f{\"u}r Mathematik und Physik, Albert-Ludwigs-Universit{\"a}t, Freiburg, Germany\\
$^{49}$ Section de Physique, Universit{\'e} de Gen{\`e}ve, Geneva, Switzerland\\
% ...
$^{89}$ Department of Physics and Astronomy, Michigan State University, East Lansing MI, United States of America\\
% ...
$^{106}$ Nikhef National Institute for Subatomic Physics and University of Amsterdam, Amsterdam, Netherlands\\
% ...
$^{112}$ Homer L. Dodge Department of Physics and Astronomy, University of Oklahoma, Norman OK, United States of America\\
$^{113}$ Department of Physics, Oklahoma State University, Stillwater OK, United States of America\\
% ...
$^{116}$ LAL, Universit{\'e} Paris-Sud and CNRS/IN2P3, Orsay, France\\
% ...
\end{flushleft}
